\documentclass[superscriptaddress,prl,twocolumn,nofootinbib]{revtex4}
\usepackage{amsmath,amssymb}
\usepackage{graphicx,epsfig}
\def\simgt{\mathrel{\lower2.5pt\vbox{\lineskip=0pt\baselineskip=0pt
           \hbox{$>$}\hbox{$\sim$}}}}
\def\simlt{\mathrel{\lower2.5pt\vbox{\lineskip=0pt\baselineskip=0pt
           \hbox{$<$}\hbox{$\sim$}}}}

\def\vev#1{\left\langle#1\right\rangle}
\def\stacksymbols #1#2#3#4{\def\theguybelow{#2}
    \def\vp{\lower#3pt}
    \def\sp{\baselineskip0pt\lineskip#4pt}
    \mathrel{\mathpalette\intermediary#1}}

\def\intermediary#1#2{\vp\vbox{\sp
     \everycr={}\tabskip0pt
     \halign{$\mathsurround0pt#1\hfil##\hfil$\crcr#2\crcr
              \theguybelow\crcr}}}

\def\gapproxeq{\stacksymbols{>}{\sim}{2.5}{.2}}

\def\beq{\begin{equation}}
\def\eeq#1{\label{#1}\end{equation}}
\def\eeqn{\end{equation}}
\def\beqa{\begin{eqnarray}}
\def\eeqa#1{\label{#1}\end{eqnarray}}
\def\eeqan{\end{eqnarray}}
\def\CR{\nonumber \\ }
\def\leqn#1{(\ref{#1})}
\begin{document}

\preprint{UCB/PTH-04-13}

\title{Late Time Neutrino Masses, the LSND Experiment and the Cosmic Microwave
Background}

\author{Z. Chacko, Lawrence J.~Hall, Steven J.~Oliver }
\affiliation{Department of Physics, University of California, Berkeley,
      and\\
      Theoretical Physics Group, Lawrence Berkeley National Laboratory,
      Berkeley, CA 94720, USA}

\author{Maxim Perelstein}
\affiliation{Institute for High-Energy Phenomenology, Cornell University,
Ithaca, NY 14850}

\begin{abstract}
Models with low-scale breaking of global symmetries in the neutrino 
sector provide an alternative to the seesaw mechanism for understanding why
neutrinos are light. Such models can easily incorporate light
sterile neutrinos required by the LSND experiment. Furthermore, the
constraints on the sterile neutrino properties from nucleosynthesis and 
large scale structure can be removed due to the non-conventional 
cosmological evolution of neutrino masses and densities. We present explicit,
fully realistic supersymmetric models, and discuss the characteristic 
signatures predicted in the angular distributions of the cosmic microwave 
background.
\end{abstract}

\maketitle
{\em Introduction ---}
The LSND experiment found evidence for the oscillations $\bar{\nu}_\mu
\rightarrow \bar{\nu}_e$ and $\nu_\mu \rightarrow \nu_e$ with an
oscillation probability of around $3 \times 10^{-3}$~\cite{LSND}
and a $\Delta m^2 \simgt 1 \mbox{eV}^2$. The
statistical evidence for the anti-neutrino oscillations is much
stronger than for the neutrino case, with some analyses finding a $5
\sigma$ effect~\cite{strumia}. While other experiments restrict the
regions of parameter space that could explain the LSND data,
they do not exclude the LSND result~\cite{PDG}.

The confirmation of solar and atmospheric neutrino oscillations has
led to a ``standard'' framework for neutrino masses, with three light
neutrinos resulting from the seesaw mechanism and the heaviest
right-handed neutrino not far from the scale of gauge coupling
unification~\cite{see-saw}. The LSND result conflicts with this framework, 
and, if confirmed by the Mini-BooNE experiment~\cite{boone}, would throw 
neutrino 
physics into a revolutionary phase. There are three major challenges to
incorporate the LSND result into the standard framework for neutrino masses.
A fourth light neutrino is needed with mass in the eV range, and,
given the $Z$ width, this neutrino must be sterile. Such states are
anathema to the seesaw mechanism. Secondly, neutrino oscillations in
the early universe would ensure that this fourth neutrino state is
thermally populated during the big bang nucleosynthesis (BBN),
changing the expansion rate and yielding light element abundances in
disagreement with observations~\cite{avoid}. In fact, theories with four 
neutrino species give very poor fits to the combined data of the  
LSND experiment and oscillation experiments with null results, while
fits in theories with 5 neutrinos are much better~\cite{three+two}. This 
would imply that $N_{\nu {\rm BBN}} \geq 5$, in gross disagreement with 
observations~\cite{deltaN,LSS}. Finally, the combination of large scale 
structure surveys and WMAP data~\cite{WMAP+LSS,LSS1} has led to a limit on the 
sum of the neutrino masses of about 0.7 eV, which is significantly less 
than the best fit values for the LSND neutrino masses~\cite{LSS,LSS1}. 
\if
Such difficulties have prompted radical proposals,
such as CPT violation in the neutrino spectrum \cite{my}, but even in
this case sterile states seem necessary \cite{cpt+}.
\fi

An alternative explanation for why the neutrinos are light has been 
explored recently: the scale of neutrino masses can be dictated by a low
scale $f$ of breaking of global symmetries in the neutrino 
sector~\cite{choo,majoron1}. A characteristic feature of
this mechanism is that, cosmologically, neutrinos remain massless until the 
symmetry-breaking phase transition, which occurs quite late in the history of 
the universe --- hence the name ``late time neutrino masses''. 
In this letter, we argue that, unlike the traditional seesaw framework, this 
alternative scenario can easily accommodate the sterile neutrinos required by 
the LSND experiment. Moreover, the cosmological evolution of neutrino
masses and densities in this scenario is non-standard and, as 
a result, the apparent contradiction between the parameters preferred by
LSND and cosmology can be avoided, provided that $f$ is of order 
100 keV. (We will show that such values of $f$ can arise naturally 
in supersymmetric theories.) At the same time, the scenario predicts 
potentially observable characteristic signatures in the cosmic microwave 
background (CMB) angular distributions. 

All of the above features are generic in models with low-scale breaking
of neutrino global symmetries, and can be understood without reference to a 
specific model. Let us outline the basic physical arguments.

The traditional seesaw framework relies on the gauge quantum numbers
of the neutrinos to explain their mass spectrum, and leads to a
picture where only active neutrinos are light. Global symmetries, on
the other hand, may involve sterile as well as active neutrinos, and
they may forbid both Dirac and Majorana mass terms. Neutrino masses
then appear as a result of spontaneous breaking of these symmetries,
so that, in this scenario, it is quite natural to expect light sterile
states.
    
If the symmetry breaking phase transition 
occurs {\it after} the BBN epoch, both active and sterile neutrinos are 
massless before and during nucleosynthesis. In this case, the
oscillations, that typically lead to thermal abundances for the sterile 
states in the traditional scenario, are absent. During BBN, the energy density 
of the sterile neutrinos (and of the scalars required to break the global 
symmetries) is determined by their temperature. As we show below, this 
temperature can be significantly lower than that of 
the rest of the cosmic fluid, provided that the rates of certain reactions   
are sufficiently low. This allows our models to evade the BBN constraints on 
$\Delta N_\nu$.

Remarkably, the limit from large scale structure on the sum of the neutrino 
masses is also easily avoided by the late time neutrino mass models. The 
breaking of the global 
symmetries gives rise to a set of Goldstone bosons, which are coupled
to both active and sterile neutrinos. This coupling 
is sufficiently strong for the sterile neutrinos to disappear
after they become non-relativistic, for example by decaying into an active
neutrino and a Goldstone boson.
As a result, the relic abundance of the sterile
neutrinos is low, and they do not significantly contribute to dark matter 
despite their large mass.

{\em Specific Models---}
There is a wide variety of late time neutrino mass models:
the neutrinos may be either Dirac or Majorana, the number 
of sterile neutrinos may vary, and different choices of global 
symmetries and their breaking patterns can be made. Let us present two
simple supersymmetric models incorporating the LSND neutrinos. We do
not consider models with a single sterile neutrino, as they are
disfavored by oscillation data. 
For concreteness we construct models with three 
mass eigenstates that are predominantly sterile. We take the global  
symmetry to be  $U(1) \times U(1)$; a simple possibility that allows a 
heavy neutrino to decay to light neutrino and a Goldstone boson. 

Our first theory has three right-handed neutrino superfields, $n$. There is 
no overall lepton number symmetry, leading to six physical Majorana neutrinos. 
Above the weak scale the theory is described by the superpotential
\beqa
W^M &=& W_{NMSSM} + W_{\nu}^M, \CR
W_{\nu}^M &=&
\lambda_{i j} l_i n_{j} h \;  \frac{\phi}{M}
+ \frac{\kappa}{3} \phi^3 + \tilde{\lambda}_{i j} n_{i} n_{j} s \;
\frac{\tilde{\phi}}{M}  + \frac{\tilde{\kappa}}{3} {\tilde{\phi}}^3,
\eeqa{super1}
where $W_{NMSSM}$ is the superpotential of the NMSSM; $\lambda$, 
$\tilde{\lambda}$, $\kappa$ and  $\tilde{\kappa}$ are coupling constants;
and the flavor indices $i$ and $j$ run from $1 \rightarrow 3$. The superfields
$l$, $h$ are the lepton and Higgs doublets of the MSSM, $s$ is the  
electroweak singlet field of the NMSSM, and $\phi$, $\tilde{\phi}$ are the
extra electroweak singlet fields whose vacuum expectation values (vevs) give 
masses to neutrinos. The non-renormalizable operators in~\leqn{super1} are 
generated by integrating out physics at scale $M$; phenomenological 
constraints discussed below imply $M\sim 10^9$ GeV.   
In theories without an $s$ field, the third operator in $W_{\nu}^M$ would be 
absent, and
we would expect three light Dirac neutrinos. (If $nn\tilde{\phi}$
were allowed, the sterile states would be much heavier than the active
states.) However, for theories such as the NMSSM, where the $s$ field
acquires a vev of order the electroweak 
scale, the Dirac and Majorana mass terms are of the same order of magnitude, 
$vf/M$, explaining
why the LSND neutrinos are quite close in mass to the active neutrinos.
$W^M$ is the most general superpotential in the neutrino
sector up to dimension four under the following discrete symmetries: 
$Z_3$, under which  all the fields except $\phi$ and  $\tilde{\phi}$ 
have charge $2 \pi/3$;
$Z'_3$, under which $s, h$ and $\bar{h}$ are uncharged, $q,l,n$ and $\phi, 
\tilde{\phi}$ have charge $2 \pi/3$, while $u^c,d^c$ and $e^c$ have 
charge $-2 \pi/3$; and $Z''_3$, under which $n$ and $\tilde{\phi}$ 
both have charge $2\pi/3$ while $\phi$ has charge $-2\pi/3$.

                  
Below the weak scale the  renormalizable effective Lagrangian for the
neutrino sector of the theory is
\begin{equation} 
\mathcal{L}^M_{\nu} =  g_{i j} \nu_i n_{j} \phi + 
\tilde{g}_{i j} n_{i} n_{j} \tilde{\phi} + {\rm{h.c.}}  + 
V(\phi, \tilde{\phi}),
\label{eq:LM}
\end{equation}
where $g = \vev{h} \lambda /M$, $\tilde{g} = \vev{s}\tilde{\lambda} /\tilde{M}$
and the scalar potential is $V = -
\mu^2 |\phi|^2 + \kappa^2 |\phi|^4 - \tilde{\mu}^2 |\tilde{\phi}|^2 +
\tilde{\kappa}^2 |\tilde{\phi}|^4$. (We have assumed that SUSY breaking 
effects generate {\em negative} soft mass$^2$ terms for $\phi, \tilde{\phi}$.)
This theory has two accidental $U(1)$ global
symmetries: one under which $\phi$ and $\nu$ are charged and another one
under which $\tilde{\phi}$, $\nu$ and $n$ are charged. (While these symmetries 
are not exact even at the renormalizable level, the terms that do not respect 
them are quite small: for example, the $\phi^3$ term in $V(\phi)$ is only 
generated at three loops and can be neglected in the present context.) 
When $\phi$ and
$\tilde{\phi}$ acquire vevs, these symmetries are broken leading to two
pseudo-Goldstone bosons $G$ and $\tilde{G}$, and giving the neutrinos a mass.

With only minor changes we can construct a theory where the six
neutrinos are Dirac. There are now three singlet left-handed sterile
neutrinos, $\nu^s_i$, and a total of 6 right-handed neutrinos
$n_\alpha$, coupled via
\beqa
W^D &=& W_{NMSSM} + {W}^D_{\nu}, \CR
W^D_{\nu} &=&
\lambda_{i \alpha} l_i n_{\alpha} h \; \frac{\phi}{M}
+ \frac{\kappa}{3} \phi^3 
+ \tilde{\lambda}_{i \alpha} \nu^s_{i} n_{\alpha} s \;
\frac{\tilde{\phi}}{M}  + \frac{\tilde{\kappa}}{3} {\tilde{\phi}}^3.
\eeqa{super2}
The superpotential $W^D$ is the most
general up to dimension four that is invariant under $Z_3 \times
Z_3' \times Z_3''$ (with $\nu^s$, like $n$, having charges $2 \pi /3$
under each $Z_3$) together with a lepton number symmetry under which
$l$ and $\nu^s$ have the same charge and $n$ the opposite charge.

Below the weak scale the renormalizable effective Lagrangian for the
neutrino sector of this theory is
\begin{equation}
\mathcal{L}^D_{\nu} = g_{i \alpha} \nu_i n_{\alpha} \phi   +
\tilde{g}_{i \alpha} \nu^s_{i} n_{\alpha} \tilde{\phi} + {\rm{h.c.}}
  + V(\phi, \tilde{\phi}).
\label{eq:LD}
\end{equation}
The theory has two approximate global symmetries: one under which
$\phi$ and $\nu$ are charged, and another under which $\tilde{\phi}$ 
and $\nu^s$ are charged. Again, $\phi$ and $\tilde{\phi}$
vevs break these symmetries leading to two pseudo-Goldstone bosons
and Dirac masses for neutrinos.

Eqs.~(\ref{eq:LM}) and~(\ref{eq:LD}) imply that $g$ and
$\tilde{g}$ are of order $m_{\nu}/f$, where $m_{\nu}$ is a scale of
order the neutrino masses, and $f$ is the scale at which the
global symmetries are broken. It is important to note that
in general the couplings of the Goldstones to the neutrinos are not
diagonal in the neutrino mass basis.  Instead, denoting the mass
eigenstates by primes, these couplings are of the form
$(g_{\alpha \beta} \nu'_\alpha n'_\beta G + \tilde{g}_{\alpha \beta}
\nu'_\alpha n'_\beta \tilde{G})$ (Dirac case) and 
$(g_{\alpha \beta} \nu'_\alpha \nu'_\beta G + \tilde{g}_{\alpha \beta} 
\nu'_\alpha \nu'_\beta \tilde{G})$ (Majorana case).

These theories provide concrete examples of a very rich set of
theories. A particularly simple theory is obtained by deleting the
$\tilde{\phi}$ field and its interactions, and removing the $Z_3''$
symmetry so that $\phi$ can couple to both doublet and singlet
neutrino mass operators. In this case there is a single flavor
diagonal $U(1)$ symmetry and hence a single Goldstone, having
diagonal couplings to neutrinos in the mass basis.

{\em Constraints---} Significant constraints on the parameter space of these
theories follow from the requirement that the total energy density in
radiation at the time of BBN does not differ significantly from the
Standard Model prediction. 
\if
In the parameter range of interest, the ``hidden
sector'' fields ($\phi$, $\tilde{\phi}$, $n$ and possibly $\nu^s$) 
are in thermal equilibrium with the ``visible sector'' fields ($\nu, \gamma, 
\ldots$) at BBN through decays and inverse decays of $\phi$: 
$\phi \leftrightarrow \nu n$ 
This, however, does not automatically mean that the energy density per degree
of freedom is the same in the hidden and visible sector. Indeed, suppose
that the visible sector is 
reheated at some point before BBN by the decoupling of heavy particles 
($\mu$, $\pi$, $\ldots$) or a phase transition. The reaction~$\nu n
\rightarrow\phi$ 
will reheat the hidden sector; however, it cannot change the total
number density in that sector, since it conserves the number of hidden
sector particles. If all reactions that {\em do not} conserve this
number are ``frozen'' ($\Gamma<H$), a chemical potential will develop
in the hidden sector, suppressing its energy density. Note that this
requirement would be difficult to satisfy in a model with a single
$\phi$ field, due to the possibility of the decay $\phi\to nn$.  
\fi
This requires that the ``hidden sector'' fields ($\phi$,
$\tilde{\phi}$, $n$ and possibly $\nu^s$, as well as the fermionic partners
of $\phi$ and $\tilde{\phi}$ which will turn out to be quite light) not be 
in thermal equilibrium with the ``visible sector'' fields ($\nu, \gamma, 
\ldots$) before and during the BBN. More precisely, we require that the two
sectors decouple at a certain temperature $T_0>$ 1 GeV, and do not recouple 
until the temperature of the visible sector drops below $T_W\sim 1$ MeV, the 
temperature at which the weak interactions decouple. If this is the case, 
the reheating of the visible sector by the decoupling of heavy particles 
($\mu$, $\pi$, $\ldots$) and possibly by the QCD phase transition will not
affect the hidden sector. Defining $r$ as the ratio of temperatures of the 
two sectors at the time of BBN, we conclude that the energy density in the 
hidden sector is suppressed by a factor of $r^4$ compared to the naive 
estimate, and $r\simlt 0.3$ allows one to avoid the BBN constraint even 
for a very large hidden sector.


The reactions that could recouple the two sectors include
a $1 \leftrightarrow 2$ process $\phi\leftrightarrow \nu n$, $2
\leftrightarrow 2$ processes such as $\nu \bar{\nu} \leftrightarrow n
\bar{n}$ and $\nu n\leftrightarrow \phi\tilde{\phi}$, $2 \leftrightarrow 3$ 
processes such as $\nu n \leftrightarrow 3\phi$, etc.  Requiring that all 
these processes be ``frozen'' ($\Gamma<H$) prior to the weak interactions 
decoupling results in the following constraints on the couplings:
\beqa
g_{ij},~~g_{i\alpha} \simlt 10^{-5},~~&~&~~g_{ij}\kappa,~~g_{i\alpha}\kappa
 \simlt 10^{-10} r^{-1}, \CR 
g_{ij} \tilde{g}_{ij},~~g_{i\alpha} \tilde{g}_{i\alpha} &\simlt& 10^{-10}
r^{-3/2}.
\eeqa{constraints}
Note that the coupling $\tilde{\kappa}$ is unconstrained.

\if
Another set of constraints comes from supernova dynamics. 
The presence 
of additional neutrino interactions can affect a supernova in two distinct
ways. The decays of $\nu_e$ into $\phi$ fields can
deleptonize the core prior to the `bounce' preventing the bounce from
taking place. Further, after the bounce, the supernova can lose energy
too rapidly through neutrino or antineutrino decays into Goldstone
fields which then free-stream out. These considerations have been used to
put bounds on the Goldstone couplings to the various neutrino species.  
While these constraints are quite model-dependent, 
they are typically at the level of $g_{ij},g_{i\alpha} 
\simlt 10^{-5}--10^{-6}$~\cite{SN}, similar to Eq.~\leqn{constraints}.
\fi

The upper bounds on the coupling $g$ can be translated into 
a lower bound on the symmetry breaking scale $f$. To interpret the LSND 
result in the model with Majorana sterile neutrinos, the low-energy theory 
must 
possess a mass term of the form $m_{ij}\nu_i n_j$, with at least some 
elements of $m$ as large as $0.1$ eV. This implies that $g_{ij}f \sim 0.1$ 
eV, and for a generic flavor structure we obtain a bound $f\gapproxeq 10$ keV.
A similar bound can be obtained for the Dirac sterile neutrino case.  

To avoid producing sterile neutrinos by oscillations prior to weak
interactions decoupling, we require that the mass terms mixing active and
sterile neutrinos not be generated until the temperature of the visible 
sector drops below $T_W$. 
Scattering in the plasma generates ``thermal'' masses for the $\phi$
bosons, $m^2(\phi)\sim \kappa^2 n_\phi(T')/T'$, 
where $T'$ is the temperature of the hidden sector. The symmetry breaking 
phase transitions for $\phi$ occurs when $m^2(\phi)\sim\mu^2$.  Using
$\mu=f\kappa$, we conclude that the temperature of the visible sector 
at the time of this phase transition is $\sim f/r$, implying that
$f\simlt r$ MeV is necessary for the success of BBN. This in turn imposes a
{\em lower} limit on the couplings, $
g_{ij}\gapproxeq r^{-1}\,10^{-7}.$ 


 
To summarize, BBN 
considerations lead to a range of the allowed values of the scale $f$,
\begin{equation}
10 \; \mbox{keV} \simlt f \simlt r \;
\mbox{MeV}. 
\label{eq:f}
\end{equation}
Considerations of the supernova dynamics may slightly raise the lower
bound; however, these constraints are strongly model dependent~\cite{SN}.
Even though the allowed values of $f$ are much lower than
the weak scale, the theory naturally allows for symmetry breaking in
this range. The only assumption necessary is that $\phi$ only feels 
supersymmetry breaking through its
couplings to $l$ and $n$. Then $\mu^2$ is of
order $g^2 m^2_{{\rm SUSY}}/ 16 \pi^2$, where $m_{{\rm SUSY}}$ is a
typical soft supersymmetry breaking mass. Since $g$ is of order
$m_{\nu}/f$ and $f$ itself is of order $\mu/\kappa$, by eliminating
$\mu$ and $g$ in favor of $f$ and $m_{\nu}$ we are led to the
expression
\begin{equation}
f \approx \sqrt{\frac{m_{\nu} \; m_{{\rm SUSY}}}{ 4 \pi \kappa}}
\end{equation}
For reasonable values of the parameters $m_{\nu} \approx$ 0.1 eV,
$m_{{\rm SUSY}} \approx$ 100 GeV, $\kappa \approx 10^{-4}$, this
yields a value of $f$ of order 3 MeV, which is quite close to
the desired range.  Analogous considerations apply to the
second symmetry breaking scale, $\tilde{f}$.

Interestingly, the large-scale structure limit on the sum of neutrino 
masses~\cite{LSS} is {\em automatically} avoided in the models discussed 
here, and does not lead to additional constraints on $f$. 
The lower bound on $g_{ij}$ obtained above
implies that the reactions $\nu \bar{\nu} \leftrightarrow n 
\bar{n}, \phi\bar{\phi}$ become unfrozen {\em before} the sterile 
neutrinos become non-relativistic. 
These reactions thermalize the hidden sector fields with the active
neutrinos. The density of thermal, sterile neutrinos of mass $m_s$ at
temperatures $T<m_s$ is suppressed by a Boltzmann factor $e^{-m_s/T}$;
the excess neutrinos disappear either via a decay process
$n\rightarrow \nu\phi$, or via an annihilation process $n\bar{n}
\rightarrow \nu\bar{\nu}$. As a result, the massive sterile neutrinos
do not make a significant contribution to dark matter. It is only the
sum of the masses of active, stable neutrinos and the Goldstones
that has to satisfy the constraints of Ref.~\cite{WMAP+LSS}.  

{\em Signals in the CMB---} The non-standard evolution of neutrino masses
and densities in our scenario leaves an imprint in the CMB inhomogeneities. 
There are two distinct, potentially observable 
effects~\cite{choo}. First, the total relativistic energy density at the 
time of last scatter is modified due to the decay and annihilation of the 
sterile neutrinos. Second, unlike in the standard cosmology, free-streaming
of the active neutrinos may be prevented by their interactions with the 
Goldstone bosons. Let us consider each of these effects.

At the time of BBN, the energy density in the hidden sector is
suppressed. 
When the reactions $\nu \bar{\nu} \leftrightarrow n \bar{n},
\phi\bar{\phi}$, $\nu n \leftrightarrow \phi$ become unfrozen, the two sectors
thermalize and 
the energy density per degree of freedom in the hidden sector
approaches that of the active neutrinos. (The active neutrinos
themselves are by this time decoupled from electrons and photons.)
These reactions, however, do not change the total relativistic energy
density: they merely transfer part of the energy from the active
neutrinos to the hidden sector states. In contrast, when the sterile
neutrinos become non-relativistic ($T\sim m_s \sim 1$ eV) and are
depleted by decays and annihilations, the total relativistic energy 
is increased: the depletion
process occurs at constant entropy, resulting in an increase in
temperature as the number of relativistic degrees of freedom
decreases.  Since the depletion of $n$'s occurs before the
matter-radiation equality, this will result in a non-standard value of
the relativistic energy density implied by the CMB measurements.
In terms of the ``effective'' number of neutrinos 
$N_{\nu,{\rm CMB}}$~\cite{choo}, our scenario predicts
\beq
N_{\nu,{\rm CMB}} = 3\left(1+\frac{n_s + 2.75n_h/g_\nu}{3+n_G/g_\nu}
\right)^{1/3}.
\eeq{Nnueff} 
Here, $g_\nu$ equals $7/4$ for Majorana neutrinos and $7/2$ for the Dirac case;
$n_h$ is the number of the ``Higgs'' (massive) components of the scalar
fields responsible for global symmetry breaking that are light enough to be
relativistic when the reactions $\nu\nu\rightarrow\phi\phi$, $\nu n
\rightarrow \phi\tilde{\phi}$ become unfrozen;
$n_s$ is the number of sterile neutrino species, and $n_G$ is the number of
Goldstone modes. (Eq.~\leqn{Nnueff} includes the contribution from the 
superpartners of the symmetry breaking scalar fields.)
For example, in the explicit models presented above,   
$n_s=3$ and $n_G=n_h=2$. 
Some typical values for $N_{\nu,{\rm CMB}}$ 
are presented in Table~\ref{table:Nnu}. For comparison,
while the current sensitivity on $N_{\nu,{\rm CMB}}$ from the WMAP and
other CMB analyses~\cite{deltaNu} is about $\pm 5$, the sensitivity of the 
Planck experiment is expected to reach the $\pm 0.20$ level, providing a 
test of our predictions.
    
 \begin{table}
\begin{center}
\begin{tabular}{|c||c|c|c||c|c|c|} \hline
    & \multicolumn{3}{|c||}{Dirac} & \multicolumn{3}{|c|}{Majorana} \\
    $n_G$ & \multicolumn{3}{|c||}{$n_s$} & \multicolumn{3}{|c|}{$n_s$} \\
 \hline 
     & 1 & 2 & 3 & 1 & 2 & 3 \\ \hline
    2& 3.59 & 3.78 & 3.95 & 3.78 & 3.92 & 4.06  \\ \hline
    3& 3.70 & 3.86 & 4.01 & 3.91 & 4.03 & 4.14  \\ \hline
    8& 4.00 & 4.11 & 4.21 & 4.22 & 4.29 & 4.35  \\ \hline
\end{tabular}
\caption{Effective number of neutrino species during the 
recombination era, $N_{\nu,{\rm CMB}}$, as determined by 
the relativistic energy density.}
\label{table:Nnu}
\end{center}
\end{table}

Furthermore, at the time of last scatter the mean free paths of the 
light neutrinos and the Goldstones are well below the Hubble scale due to
the process $\nu_i \leftrightarrow \nu_j G$. 
The absence of free-streaming leads to a shift in the positions 
of the CMB peaks at large $l$~\cite{Shinsky,choo}.  
This shift (relative to the Standard Model prediction) is given by
 \beq 
   \Delta l_n = 23.3 - 13.1 \left( \frac{g_\nu (3-n_S)}{(3g_\nu + n_G)
                (1/N_{\nu, {\rm CMB}}+.23)} \right)
 \eeq{deltaL} 
where $n_S$ is the number of light neutrinos that are scattering during the 
eV era. (It is possible that $n_S<3$ if one of the neutrinos is
massless or very nearly so, or if $m_G$ is large enough to make the
process $\nu_i \rightarrow \nu_j G$ kinematically forbidden for some flavors).
Eq.~\leqn{deltaL} provides another experimentally testable prediction of 
our scenario. 

In the theory with no $\tilde{\phi}$ and a single Goldstone, the
neutrino decays are absent so that scattering can only occur via the
$2 \leftrightarrow 2$ processes $\nu \nu \leftrightarrow GG$ and $\nu
G \leftrightarrow \nu G$. In this case the number of neutrino species
which scatter is very sensitive to $f$ and to whether the neutrino
spectrum is hierarchical, inverted or degenerate. 

{\em Acknowledgments---} M.~P. is grateful to Alexander Friedland for 
useful discussions related to this work.
Z.~C., L.~H. and S.~O. are supported in part by 
the U.S. Department of Energy under Contract DE-AC03-76SF00098 and
DE-FG03-91ER-40676, and in part by the NSF grant
PHY-00-98840. M.~P. is supported by the NSF grant PHY-0098631.  While
completing this work the authors became aware of \cite{Beacom:2004yd},
where the bound on $\Sigma m_{\nu,i}$ is avoided by having heavy
neutrinos annihilate to light scalars; but only at the expense of
increasing $N_{\nu,BBN}$ significantly above 3.

\end{document}